\documentclass[aip,chaos,amsmath,amssymb,reprint,preprintnumbers]{revtex4-1}

\usepackage{graphicx}
\usepackage{dcolumn}
\usepackage{bm}

\newcommand{\be}{\begin{equation}}
\newcommand{\ee}{\end{equation}}

\usepackage{color}
\newcounter{commentzaehler}
\begin{document}
\preprint{Chaos, in print, Dec. 2011}
\title{Adaptive Tuning of Feedback Gain in Time-Delayed Feedback Control}
\author{J.~Lehnert}

\affiliation{Institut f\"{u}r Theoretische Physik, TU Berlin,
Hardenbergstra{\ss}e 36, D-10623 Berlin, Germany}
\author{P.~H\"ovel}
\affiliation{Institut f\"{u}r Theoretische Physik, TU Berlin,
Hardenbergstra{\ss}e 36, D-10623 Berlin, Germany}
\affiliation{Bernstein Center for Computational Neuroscience, Humboldt-Universit\"at zu Berlin
Philippstr. 13, 10115 Berlin, Germany}
\author{V.~Flunkert}
\affiliation{Institut f\"{u}r Theoretische Physik, TU Berlin,
Hardenbergstra{\ss}e 36, D-10623 Berlin, Germany}
\author{P.~Yu.~Guzenko}
\affiliation{SPb State Polytechnical University,
Politechnicheskaya str., 29, St.Petersburg,  195251, Russia }
\author{A.~L.~Fradkov}
\affiliation{ Institute for Problems
of Mechanical Engineering, Russian Academy of Sciences,
Bolshoy~Ave,~61, V.~O.,~St.~Petersburg, 199178~Russia }
\affiliation{SPb State University, Universitetskii pr.28,
  St.Petersburg, 198504 Russia}
\author{E.~Sch\"oll}
\affiliation{Institut f\"{u}r Theoretische Physik, TU Berlin,
Hardenbergstra{\ss}e 36, D-10623 Berlin, Germany}
\date{\today}

\begin{abstract}
We demonstrate that time-delayed feedback control can be improved by
adaptively tuning the feedback gain. This adaptive controller is applied to the 
stabilization of an unstable fixed point and an unstable periodic orbit embedded in 
a chaotic attractor. The adaptation algorithm is constructed using the speed-gradient
method of control theory. Our computer simulations show that the adaptation
algorithm can find an appropriate value of the feedback gain for
single and multiple delays. Furthermore, we show that our method is robust to noise and
different initial conditions.
\end{abstract}
\maketitle

\begin{quotation}
  {\bf The control of nonlinear systems is a central topic in
    dynamical system theory, with a diverse range of
    applications. Adaptive control schemes have emerged as a new type
    of control method that optimizes the control parameters with
    respect to an appropriate goal function, thereby minimizing, for
    instance, the consumed power or the time needed to reach the
    control goal.  In this work we combine time-delayed feedback
    control, an established method from chaos control, with an
    adaptive speed-gradient scheme to optimize the control force.  We
    demonstrate how this combined scheme can be utilized to stabilize
    various target states, e.g., unstable fixed points or periodic orbits,
    with little or no apriori knowledge about the target state.  We
    also investigate the robustness of the method to noise and perturbations.  }

\end{quotation}

\section{Introduction}
Stabilization of unstable and chaotic systems forms an important field
of research in nonlinear dynamics. A variety of control
schemes have been developed to control periodic orbits as well as
steady states \cite{OTT90,SCH07}. A simple and efficient scheme,
introduced by Pyragas \cite{PYR92}, is known as \textit{time-delay
  autosynchronization} (TDAS). This control method generates a
feedback from the difference of the current state of a system to its
counterpart some time units $\tau$ in the past. Thus, the control
scheme does not rely on a reference system and has only a small number
of control parameters, i.e., the feedback gain $K$ and time delay
$\tau$. It has been shown that TDAS can stabilize both unstable
periodic orbits, e.g., embedded in a strange attractor
\cite{PYR92,BAL05} as well as unstable steady states
\cite{AHL04,ROS04a,HOE05}. In the first case, TDAS is most efficient
and noninvasive if $\tau$ corresponds to an integer multiple of the
minimal period of the orbit. In the latter case, the method works best
if the time delay is related to an intrinsic characteristic timescale
given by the imaginary part of the system's eigenvalue \cite{HOE05}. A
generalization of the original Pyragas scheme, suggested by Socolar et
al. \cite{SOC94}, uses multiple time delays. This \textit{extended
  time-delay autosynchronization} (ETDAS) introduces a memory
parameter $R$, which serves as a weight of states further in the past.
In Ref.~\onlinecite{DAH07} it is shown that, this method is able to control
an unstable fixed points for a larger range of parameters compared to
the original TDAS scheme.  A variety of analytic results about
time-delayed feedback control are known
\cite{BLE96,JUS97,JUS98,PYR01}, for instance, in the case of long time
delays \cite{YAN06}, transient behavior \cite{HIN11}, unstable
spatio-temporal patterns \cite{BAB02}, or regarding the odd number limitation
\cite{NAK97}, which was refuted in Refs.~\onlinecite{FIE07,JUS07}.

In the present paper, we apply the speed-gradient method
\cite{FRA79, FRA98a,FRA05b,FRA07,GUZ08} to adaptively tune the
feedback gain $K$, which is used in both TDAS and ETDAS control methods, and utilize this
scheme to stabilize an unstable focus in a generic model, and an
unstable periodic orbit embedded in a chaotic attractor. The former
model is the generic linearization of a system with an unstable fixed
point close to a Hopf bifurcation. The speed-gradient method is a
well known adaptive control technique that minimizes a predefined goal
function by changing an accessible system parameter appropriately. The
adaptation of the feedback gain may be useful, in particular, for
systems with slowly changing parameters or when the domain of
stability is unknown. There are several other approaches to adaptive control of nonlinear systems in the control literature \cite{FRA99a, AST08a, KRS95}. 
Here we have chosen the speed-gradient method because it is simple and robust.

This paper is organized as follows: In Sec.~\ref{sec:ufp}, we develop
the adaptation algorithm using the example of an unstable focus. In
Sec.~\ref{sec:upo}, we apply the adaptive control scheme to stabilize
an unstable periodic orbit embedded in the chaotic attractor of the
R\"ossler system. Finally, we conclude with Sec.~\ref{sec:conclusion}.

\section{\label{sec:ufp} Stabilization of an unstable fixed point}
First, we will consider stabilization of an unstable
fixed point by time-delayed feedback. Unlike in previous works (see e.g. \cite{HOE05,DAH07}
and references therein), we do not fix the feedback gain a priori, but
tune it adaptively. We consider a general dynamical system given
by a nonlinear vector field ${\bf f}$:
\begin{eqnarray}
        \label{eqn:dyn0}
                \dot{\bf X}(t) = {\bf f}[{\bf X}(t)]
\end{eqnarray}
with ${\bf X} \in \mathbb{R}^n$ and an unstable fixed point ${\bf
  X}^*$ solving ${\bf f}({\bf X^*})=0$. The stability of this fixed point is obtained by
linearizing the vector field around ${\bf X}^*$. Without loss of
generality, let us assume ${\bf X}^*=0$. In the following we will
consider the generic case of a two-dimensional unstable focus, i.e., a
system close to a Hopf bifurcation, for which the linearized equations
can be written in center manifold coordinates $x, y \in  \mathbb{R}$ as follows:
\begin{subequations} 
\label{eqn:dyn1}
\begin{eqnarray}
        \dot{x} &=& \lambda \, x + \omega \, y \\
        \dot{y} &=& -\omega \, x + \lambda \, y,
\end{eqnarray}
\end{subequations}
where $\lambda$ and $\omega$ are positive real numbers. $\lambda$ may
be viewed as the bifurcation parameter governing the distance from the
instability threshold, i.e., a Hopf bifurcation, and $\omega$ is
the intrinsic eigenfrequency of the focus. For notational
convenience, Eq.~(\ref{eqn:dyn1}) can be  rewritten as
\begin{eqnarray}
        \label{eqn:dyn2}
        \dot{\bf X}(t) = {\bf A} \; {\bf X}(t).
\end{eqnarray}
The eigenvalues $\Lambda_0$ of the $2 \times 2$ matrix  ${\bf A}$ are given by
$\Lambda_0 = \lambda \pm i \omega$, so that for $\lambda > 0$ and
$\omega \neq 0$ the fixed point is an unstable focus. We now
apply time-delayed feedback control \cite{PYR92} in order to stabilize this fixed point:
\begin{subequations}
\label{eqn:uss}
\begin{eqnarray}
        \dot{x}(t) &=& \lambda \, x(t) + \omega \, y(t) - K [x(t) - x(t-\tau)]\\
        \dot{y}(t) &=& -\omega \, x(t) + \lambda \, y(t) - K [y(t) - y(t-\tau)],
\end{eqnarray}
\end{subequations}
where the feedback gain $K$ and the time delay $\tau$ are real
numbers. We assume that the value of $\tau$ is known and appropriately
chosen.  Mathematically speaking, the goal of the control method is to
change the sign of the real part of the eigenvalue, leading to a decay
of perturbations from the target fixed point.

Since the control force applied to the $i$th component of the
system involves only the same component, this control scheme is
called diagonal coupling \cite{BEC02} and is suitable for an
analytical treatment. Note that the feedback term vanishes if the
fixed point is stabilized since $x^*(t-\tau)=x^*(t)$ and
$y^*(t-\tau)=y^*(t)$ for all $t$, indicating the noninvasiveness
of the TDAS method.

To obtain an adaptation algorithm for the feedback gain $K$ according to
the standard procedure of the speed-gradient method
\cite{FRA98a,FRA05b,FRA07,GUZ08,AST08}, let us choose the goal function or
cost function as follows:
\begin{equation}
\label{eq:cost}
Q({\bf X})=\frac{1}{2}\left\{[x(t) - x(t-\tau)]^2+[y(t) - y(t-\tau)]^2\right\}. 
\end{equation}
Successful control yields $Q({\bf X}(t))\rightarrow 0$ as $t \rightarrow \infty$. The speed-gradient algorithm in the differential form is
given by $\dot K = -\gamma \nabla_{K} \dot Q$, where $\gamma>0$ is
the adaptation gain and $\nabla_K$ denotes $\partial / \partial K$ . Thus, we need to calculate the gradient -- with
respect to the feedback gain $K$ -- of the rate of change of the cost
function.
For the above cost function Eq.~\eqref{eq:cost} we obtain:
\begin{widetext}
 \begin{eqnarray}
    \label{eqn:sgx}
        \dot{Q} &=& [x(t) - x(t-\tau)] [\dot x(t) - \dot x(t-\tau)] +   [y(t) - y(t-\tau))][\dot y(t) - \dot
y(t-\tau)].
\end{eqnarray}

The time derivatives of $x$ and $y$ are given by Eqs.(\ref{eqn:uss}). Thus, the speed-gradient method leads to
the following equation for the feedback gain:
\begin{eqnarray}
    \label{eq:sgxalg}
        \dot K(t) &=& \gamma \{[x(t)-x(t-\tau)][x(t)-2x(t-\tau)+x(t-2\tau)] 
           +[y(t)-y(t-\tau)][y(t)-2y(t-\tau)+y(t-2\tau)]\}.
\end{eqnarray}
\end{widetext}

\begin{figure}[th!] 
\centering
\includegraphics [width= \linewidth,angle=0]{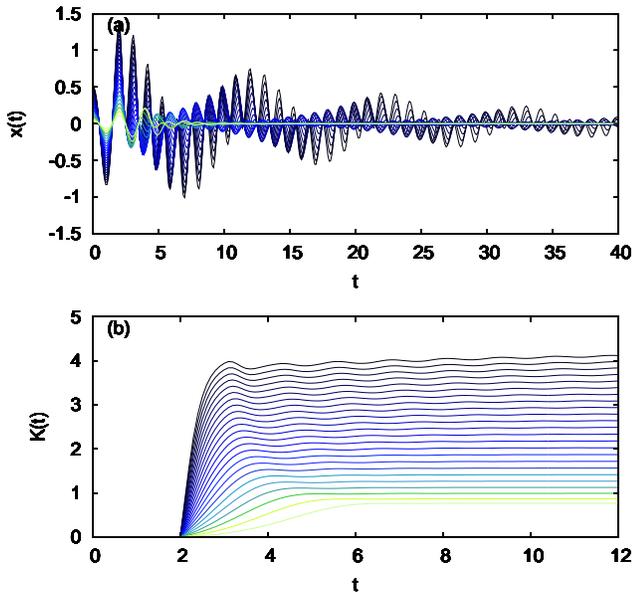}
\caption {(Color online) Adaptive control of the fixed point: (a) Time
  series $x (t)$ and (b)  feedback gain $K(t)$  for different initial conditions:
$x(0) \in [0.02,0.5]$ in steps of 0.02 (from light (green) to dark
(blue); in panel (b) from top to bottom), $y(0)=0$. 
Parameters: $\lambda=0.5$, $\omega=\pi$, $\gamma=1$, $\tau=1$.
\label{ab_bild}}
\end{figure}

Owing to homogeneity the right hand sides of Eqs.~\eqref{eqn:uss} and
\eqref{eq:sgxalg}, without loss of generality the adaptation gain $\gamma$ can be chosen
as 1, because Eqs.~\eqref{eqn:uss} and \eqref{eq:sgxalg} can be
rescaled by transformation $x(t) \longrightarrow  x(t)/\sqrt{\gamma}$
and $y(t) \longrightarrow  y(t)/\sqrt{\gamma}$.

Figure~\ref{ab_bild} depicts the time series of $x$ and $K$ according
to Eqs.~(\ref{eqn:uss}) and (\ref{eq:sgxalg}) for different initial
conditions $x(0) \in [0.02,0.5]$ in steps of 0.02 from light (green)
to dark (blue) and $y(0)=0$. In all simulations  $x(t)=y(t)=0$ for
$t<0$ and $K(t)=0$ for $t\le 2\tau$.
The parameters are chosen as
$\lambda=0.5$, $\omega=\pi$, and $\tau=1$. Figure~\ref{ab_bild}(a)
shows that the adaptation algorithm works for a large range of initial
conditions. Naturally, for initial conditions close to the fixed point
the goal is reached faster.  If the system starts initially too far
from the fixed point ($x(0)>0.85$, $y(0)=0$) the control fails
(curves not shown). Note, however, that the basin of attraction can be
enlarged by increasing $\gamma$. In fact, due to the scaling,
invariance, the maximum value of $|x(0)|$ that still leads to
successful control is proportional to $\sqrt{\gamma}$.

In Ref.~\onlinecite{HOE05} it was shown that in the $(K,\tau)$-plane tongues
exist for which the fixed point can be stabilized,
i.e., for a given $\tau$ there is a
$K$-interval for which the control is successful. As can be seen
in Fig.~\ref{ab_bild}(b), the adaptive algorithm converges to some appropriate value of $K$ in this interval depending
upon the initial conditions.

\begin{figure}[th!]
\centering
\includegraphics [width= \linewidth,angle=0]{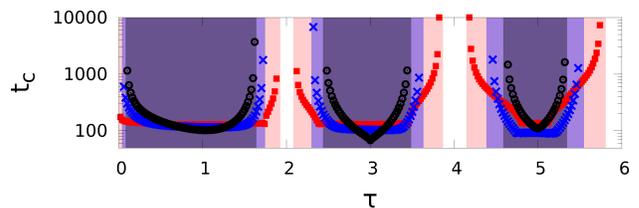}
\caption {(Color online) Transient time  $t_c$ after which the control goal is reached in
  dependence of the delay time $\tau$ for TDAS (black circles) and ETDAS with $R=0.35$ ((blue) crosses), and
  $R=0.95$ ((red) squares). The dark (dark purple), medium (bright
  purple), and light (red) shaded regions denote the
possible range of $\tau$ for $R=0, 0.35$, and $0.95$, respectively. Parameters as in Fig.~\ref{ab_bild}. \label{tau_bild}}
\end{figure}

Figure~\ref{tau_bild} demonstrates that the algorithm works for a
range of $\tau$, i.e., for any value of $\tau$ within the domain of stability of the TDAS control
\cite{HOE05}. Black empty circles depict the transient time $t_c$ after which the control
goal is reached in dependence on the time delay $\tau$. This is the
case if the cost function $Q$ becomes sufficiently small. We define
the transient time by the bound $\left<Q\right> \equiv \int_{t_c-2\tau}^{t_c}
Q(t')dt'<2\tau \times 10^{-10}$.
The dark (dark purple) shaded
regions correspond to the analytically
obtained $\tau$-intervals of the Pyragas control \cite{HOE05}. 
Inside these intervals, $t_c$ has a finite value confirming that the adaptive control scheme adjusts the feedback
gain $K$  to an appropriate value. For a comparison with the transient
time of TDAS see Ref.~\onlinecite{HIN11} where a power law scaling $t_c
\sim (K-K_c)^{-1}$ with respect to the fixed feedback gain $K$ has been
found (here $K_c$ corresponds to the boundaries of stability). 
The curves corresponding to non-zero memory parameter $R$ (crosses and
squares) will be discussed below where the speed-gradient
method is applied to the ETDAS scheme.

For a thorough analysis of the stability of the fixed point, we
perform a linear stability analysis for the system Eqs.~\eqref{eqn:uss}, \eqref{eq:sgxalg}. This system has the fixed point $(0,0,K^*)$ for any $K^*=const$.
Linearization around the fixed point and the ansatz $\delta x, \delta y, \delta K \propto\exp(\Lambda t)$ yields a
transcendental eigenvalue equation
\begin{widetext}
\begin{eqnarray}  \label{eqn:sgxydet}
0&=&\det\left[
    \begin{array}{ccc}
    \lambda - K(1-e^{-\Lambda\tau})-\Lambda & \omega & 0\\
    -\omega & \lambda- K(1-e^{-\Lambda\tau})-\Lambda & 0 \\
    0 & 0 & -\Lambda   
    \end{array}\right]\\
    &=&-\Lambda [\lambda+i\omega - K(1-e^{-\Lambda\tau})-\Lambda][\lambda-i\omega -
K(1-e^{-\Lambda\tau})-\Lambda],
\end{eqnarray}
\end{widetext}
which can be solved numerically. This equation is equal to the case of
Pyragas control with constant feedback gain considered in
Ref.~\onlinecite{HOE05} except for the factor $\Lambda$. Thus, the
adaptively controlled system has an additional eigenvalue at
$\Lambda=0$. It results from the translation invariance of the system in the
direction of $K$ on the fixed point line $(0,0,K)$. This means that
the $K$ values
found in the case of the standard Pyragas control lead again to a
stabilization of the fixed point. The advantage of an adaptive
controller is that an appropriate feedback gain is realized in an
automated way, i.e., without prior knowledge of the domain of
stability, as long as a stability domain exists for this value of
$\tau$.

\begin{figure}[th!]
\centering
\includegraphics [width= \linewidth,angle=0]{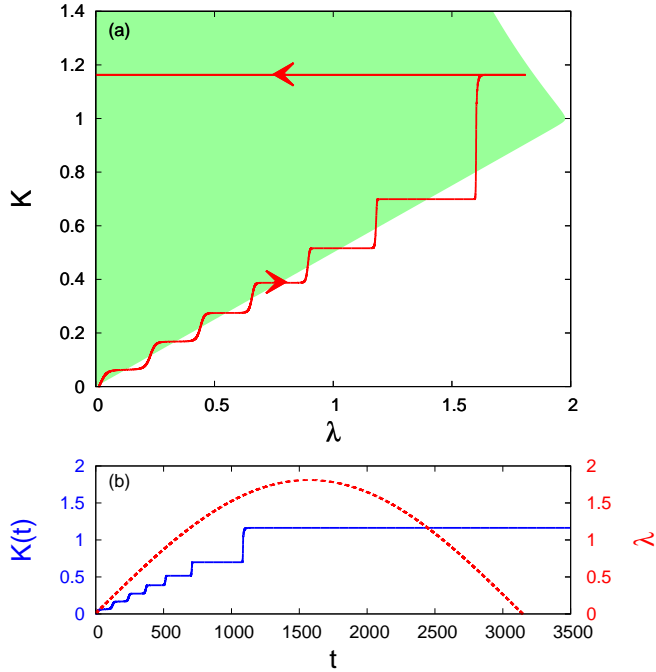}
\caption {(Color online) Adaptive control of the fixed point for slowly drifting system parameter $\lambda$. 
(a) Adaptive adjustment of $K$ in the ($\lambda,K$) plane. Green (gray) shaded region: region of stability of the standard Pyragas control. Red (solid)
line with arrow: adaptation of feedback gain $K$ if $\lambda$ is slowly changed ($\lambda(t)=0.01+1.8\sin(0.001t)$)
(b) Corresponding time series $K(t)$  (blue solid line) and $\lambda(t)$
(dashed red line). Other parameters as in
Fig.~\ref{ab_bild}. \label{lambda_K_plane}}
\end{figure}

An additional advantage of an adaptive control scheme is that it
allows one to follow slow changes of the system parameters, which are
usually present in experimental situations. To test the ability of our
adaptive control scheme to cope with such parameter drifts, we slowly vary
$\lambda$ in the following way: $\lambda(t)=0.01+1.8\sin(0.001t)$. The result is
illustrated in Fig.~\ref{lambda_K_plane}. In Fig.~\ref{lambda_K_plane}(a) the region of stability of
the standard Pyragas control in the ($\lambda,K$) plane (see
Ref.~\onlinecite{HOE05}) is marked by green (gray) shading. If now $\lambda$
is slowly increased from its initial value 0.01, $K$ follows the change in such a way that whenever the lower boundary of the stability region is crossed and the fixed point becomes unstable, the adaptation algorithm adjusts $K$ such that the stable region is re-entered.
This creates a step-like trajectory in the ($\lambda,K$) plane,
which is depicted as a red (solid) curve with an arrow. Finally, if $\lambda$ is decreased again, $K$ does not change because it already has attained a value for which the control works in a broad $\lambda$-interval resulting in a horizontal
trajectory in the ($\lambda,K$) plane. Fig.~\ref{lambda_K_plane}(b) depicts the corresponding time series of $K(t)$ as a blue (solid) curve, and of the drifting parameter $\lambda(t)$ as a red (dashed) curve, respectively.
 
To test the robustness of the control algorithm, we add Gaussian white
noise $\xi_i$ ($i=1,2$) with zero mean and unity variance ($\langle
\xi_i(t) \rangle = 0$, $\langle \xi_i(t) \xi_j(t-t') \rangle =
\delta_{ij}\delta (t-t')$) to the system variables $x$ and $y$:
\begin{subequations}
\label{dia_eqn:ussr}
\begin{eqnarray}
        \dot{x}(t) &=& \lambda \, x(t) + \omega \, y(t) \nonumber \\&& - K(t) [x(t) - x(t-\tau)] + D \xi_1(t) \\
        \dot{y}(t) &=& -\omega \, x(t) + \lambda \,y(t)\nonumber \\&&  - K(t) [y(t) - y(t-\tau)] +  D \xi_2(t),
\end{eqnarray}
\end{subequations}
where $D$ is the strength of the noise.
 
\begin{figure}[th!]  
\centering
\includegraphics[width=\linewidth]{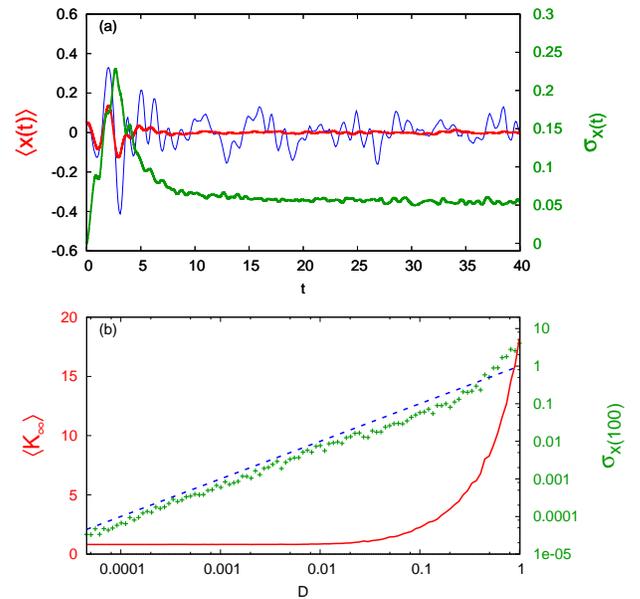}
\caption {(Color online) Robustness to noise. (a): Thick solid (red) curve: ensemble average $\langle x(t) \rangle$ of 200
realizations; thin (blue) curve: 
$x(t)$ for one example  trial; gray (green) curve: corresponding  standard
deviation $\sigma_{x(t)}$ of $x(t)$ for a fixed noise intensity
$D=0.1$. (b): (Green) crosses: standard deviation $\sigma_{x(100)}$ of $\langle
x(t=100) \rangle$; dashed (blue) line: standard deviation of the input noise
given by $D$; black (red) curve: asymptotic value $K_{\infty}$ of the feedback gain.
Parameters: $\gamma=0.001$, $x(0)=0.05$, $y(0)=0$. Other parameters as in Fig.\ref{ab_bild}.\\
\label{neu_x_average}}
\end{figure}

In Fig.~\ref{neu_x_average}(a)  the
ensemble average over 200 realizations, i.e., $\langle x(t)
\rangle=1/200 \sum^{200}_{i=1}x_i(t)$, for $D=0.1$ (intermediate
noise)  is depicted as a solid (red) curve. The thin (blue) curve exemplarily depicts one realization. The corresponding standard deviation $\sigma_{x(t)}$ of
$x(t)$ is shown as a gray (green) curve. The control is successful in
all realizations: For large $t$ the mean $\langle x(t) \rangle$
fluctuates around the fixed point value at zero, due to the finite number
of realizations. The standard deviation approaches a value smaller
than the standard deviation of the input noise.

This is further elaborated in Fig.~\ref{neu_x_average}(b), which
depicts the standard deviation $\sigma_{x}$ at $t=100$ versus the
noise strength $D$ as (green) crosses. If $D$ becomes too large the
standard deviation exceeds the one of the input noise indicated by the
dashed (blue) line. This is the case for $D \gtrsim 0.4$. Then the
control algorithm will generally fail (time series not shown here): The
oscillations of $x(t)$ become larger with increasing $t$.
Accordingly, the standard deviation $\sigma_{x(t)}$ increases with $t$
indicating that the dynamics is dominated by noise which forces
at least some of the realizations to diverge. The black (red) curve in
Fig.~\ref{neu_x_average}(b) depicts the asymptotic value $K_{\infty}$
of the feedback gain. For intermediate noise strength, an increased
feedback gain $K$ compensates the influence of noise ensuring that the
control is still successful. For too large $D$, $K$ increases to a
value beyond the domain of stability and stabilization cannot be
achieved.
 
We conclude that the adaptive algorithm is quite robust to noise (the
escape rate is vanishingly small for $D\lesssim 0.4$) and only
fails for large noise ($D \gtrsim 0.4$).  Our method allows for
finding the appropriate $K$ for all values of $\tau$ for which the standard
Pyragas control stabilizes the fixed point and is able to follow slow
drifts in the system parameters.

Note that the method still works if the control term is added only to the
$x$-component. Then, using  $Q(x)=[x(t) - x(t-\tau)]^2/2$ as a goal function leads to qualitatively very similar
results. This observation becomes relevant for experimental
realizations of the time-delayed feedback control when only
certain components of the system under control are accessible for
measurements \cite{FLU11}.

Next, we consider the ETDAS scheme \cite{SOC94}
\begin{eqnarray}
        \label{eq:dyn}
        \dot{\bf X}(t) = {\bf A} \; {\bf X}(t) - {\bf F}(t),
\end{eqnarray}
where the ETDAS control force ${\bf F}$ can be written as
\begin{subequations}
\begin{align}
    \label{eq:force}
    {\bf F}(t) &=K \sum_{n=0}^{\infty}R^{n}\left[{\bf X}(t-n\tau)-{\bf X}\textbf{(}t-(n+1)\tau\textbf{)}\right] \\
    &=K \left[{\bf X}(t)-(1-R)\sum_{n=1}^{\infty}R^{n-1}{\bf X}(t-n\tau)\right]\\
    &=K \left[{\bf X}(t)-{\bf X}(t-\tau)\right]+R {\bf F}(t-\tau).\label{eq:force_recursive}
\end{align}
\end{subequations}
Here, $R \in (-1,1)$ is a memory parameter that
takes into account those states that are delayed by more than one
time interval $\tau$. Note that $R=0$ recovers the TDAS control
scheme introduced by Pyragas \cite{PYR92}. The first form of the
control force, Eq.~(\ref{eq:force}), indicates the noninvasiveness of
the ETDAS method because ${\bf X}^* (t-\tau)={\bf X}^*(t)$ if the
fixed point is stabilized. The third form,
Eq.~(\ref{eq:force_recursive}), is suited best for an experimental
implementation since it involves states further than $\tau$ in the
past only recursively.

To apply a speed-gradient adaptation algorithm for the feedback gain $K$, we follow the same strategy as before
and choose the goal function as $Q(x)=[(x(t) - x(t-\tau))^2 + (y(t) - y(t-\tau))^2]/2$. Using again $\dot{K}=-\gamma
\nabla_K\dot{Q}$, we obtain for a diagonal control scheme

\begin{widetext}
\begin{eqnarray}
    \label{eqn:esgxyalgrec}
        \dot K(t)=\gamma \{ (x(t)-x(t-\tau))[ (x(t)-2x(t-\tau)+x(t-2\tau))+ R S_x(t-\tau)]
         \nonumber \\
        + (y(t)-y(t-\tau))[ (y(t)-2y(t-\tau)+y(t-2\tau))+ R S_y(t-\tau)]
        \}
\end{eqnarray}
with the abbreviations
\begin{eqnarray}
    \label{eqn:esgxysrec}
S_x(t)
    &=&\sum_{n=0}^{\infty}R^{n}[x(t-n\tau)-2x(t-(n+1)\tau)+x(t-(n+2)\tau)] 
    =[x(t)-2x(t-\tau)+x(t-2\tau)]+ R S_x(t-\tau) \nonumber\\
S_y(t)
    &=&\sum_{n=0}^{\infty}R^{n}[y(t-n\tau)-2y(t-(n+1)\tau)+y(t-(n+2)\tau)] 
    =[y(t)-2y(t-\tau)+y(t-2\tau)]+ R S_y(t-\tau).
\end{eqnarray}
\end{widetext}

In Ref.~\onlinecite{DAH07} the domains of stability for which ETDAS works
were obtained analytically. The intervals of $\tau$ increase with $R$ and are
larger than in the case of TDAS ($R=0$). 

Figure~\ref{tau_bild} depicts the transient time $t_c$ in dependence of $\tau$
for $R=0.35$ and $0.95$ as (blue) crosses and (red) squares, respectively. The light (red) and medium (purple) shaded
regions indicate the ranges of stability of $\tau$ \cite{DAH07}. For odd multiples of half of the intrinsic period
$T_0 \equiv2\pi/\omega$, i.e., $\tau=T_0/2(2n+1), n\in\mathbb{N}$, $t_c$ is small,
demonstrating the efficiency of the adaptive algorithm. Towards the
boundary of the domain of stability, $t_C$ increases but remains finite. The
control algorithm only fails very close to the border of the intervals
of $\tau$. We conclude that the adaptive control algorithm for
ETDAS converges to appropriate values of $K$ and stabilizes the fixed
point even for parameters where TDAS fails.

\section{\label{sec:upo} Stabilization of an unstable periodic orbit  in the R\"ossler system}

\begin{figure}[th!]
\includegraphics[width=\linewidth,angle=0]{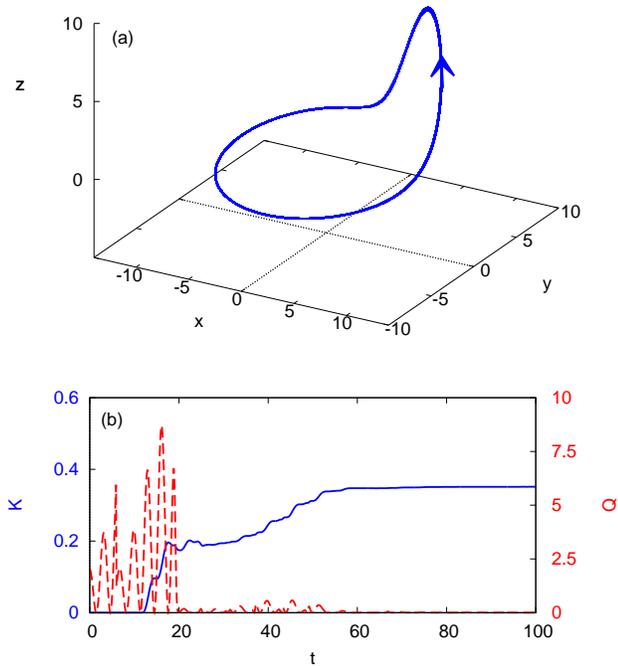}
\caption{\label{fig:ress_xyzxyxtytztKt}(Color online) Adaptive control
  of an unstable periodic orbit in the  R\"ossler attractor
  Eq.~\eqref{eqn:ress_sys}. (a): Phase
portrait (after a transient time of 150 time units). (b): Time series
of $K(t)$ with adaptive control given by
Eq.~(\ref{eq:ress_alg}) as solid (blue) curve. The dashed (red) curve shows the goal function $Q$.  Parameters:
$a=0.2$, $b=0.2$, $\mu
=6.5$, $\gamma=0.1$, $\tau=5.91679$.}
 \end{figure}

 In this section we apply the adaptive delayed feedback control
 algorithm to the
 R\"ossler system which is a paradigmatic model for chaotic
 systems. The system exhibits chaotic oscillations born via a cascade
 of period-doubling bifurcations and is given by the following
 equations including the control term:
\begin{subequations}
\label{eqn:ress_sys}
\begin{eqnarray}
        \dot{x}(t) &=& - y(t) -z(t) - K [x(t) - x(t-\tau)]\\
        \dot{y}(t) &=&  x(t) + a y(t)\\
    \dot{z}(t) &=& b + z(t) [x(t) - \mu].
\end{eqnarray}
\end{subequations}
In the following, we fix the parameter values as $a=0.2$, $b=0.2$, and
$\mu =6.5$ in the chaotic regime.
Unstable periodic orbits with periods $T_1\approx
5.91679$ ("period-1 orbit") and $T_2\approx 11.82814$ ("period-2
orbit") are embedded in the chaotic attractor.
As shown in Ref.~\onlinecite{BAL05} by a bifurcation analysis, application of the delayed feedback
of Pyragas type with $\tau =T_1$ and $0.24<K<2.3$ stabilizes the
period-1 orbit, and it becomes the only attractor of the system.
In Ref.~\onlinecite{JUS97} it was predicted analytically by a
linear expansion that control is realized only in a finite range
of the values of $K$: At the lower control boundary the limit
cycle should undergo a period-doubling bifurcation, and at the
upper boundary a Hopf bifurcation occurs generating a stable or an
unstable torus from a limit cycle (Neimark-Sacker bifurcation).

We use $Q(x)=[x(t) - x(t-\tau)]^2/2$ as a goal function and as in the
previous Section obtain the speed-gradient adaptation algorithm for $K$ \cite{GUZ08}:
\begin{equation}
    \label{eq:ress_alg}
        \dot K(t)=\gamma [x(t)-x(t-\tau)][x(t)-2x(t-\tau)+x(t-2\tau)]
\end{equation}
with the initial value $K(0)=0$. 

Figure~\ref{fig:ress_xyzxyxtytztKt}(a) depicts the time series of a
stabilized orbit for a time delay
$\tau=T_1$. Panel (b) shows that the adaptation algorithm converges
to an appropriate value of $K$ and the cost function tends to zero.

Contrary to the previous case, it is not possible to set the
adaptation gain $\gamma$ to $1$ by rescaling the system but the value of $\gamma$ is crucial for successful control. To explore the role of
$\gamma$, we determine the fraction of realizations $f_c$ where the control
goal is reached as a function of $\gamma$. The initial conditions are
Gaussian distributions with the mean $\left<x(0)\right>=\left<y(0)\right>=\left<z(0)\right>=0$, respectively, and
the standard deviations are $\sigma_{x(0)}=
\sigma_{y(0)}=\sigma_{z(0)}=1$. It is assumed that the control goal is
reached at time $t_c$ if the following condition holds:
$\left<Q\right> \equiv \int_{t_c-2\tau}^{t_c} Q(t')dt'<0.002\tau$. 

\begin{figure}[th!]
  \includegraphics[width=\linewidth,angle=0]{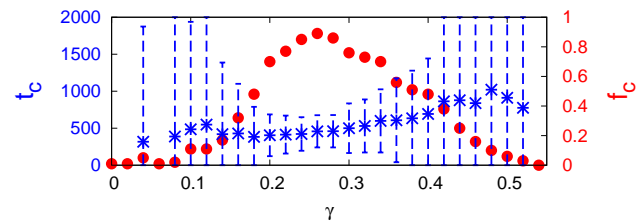}
  \caption{\label{fig:gamma}(Color online) Adaptive control of the
    R\"ossler system. Full (red) circles: fraction of
    realizations $f_c$ where the adaptive control algorithm stabilized
    the orbit versus the adaptation gain  $\gamma$; black (blue) crosses:
    Average time $t_c$ after which the control goal is reached
    versus $\gamma$; dotted (blue) lines: error bars (standard
    deviation) corresponding to $t_c$. Other parameters as in
    Fig.~\ref{fig:ress_xyzxyxtytztKt}. Total number of realizations: 100.}
\end{figure}

Figure~\ref{fig:gamma} depicts $f_c(\gamma)$ ((red) circles) and $t_c(\gamma)$
((blue) crosses) demonstrating that the optimal adaptation gain is
around $\gamma=0.26$. For $\gamma$ close to this
value, the algorithm converges fast and reliably. Accordingly, the
standard deviation of $t_c$ is small. 

This demonstrates that for appropriate values of $\gamma$, the chaotic
dynamics can  be controlled. 

\section{\label{sec:conclusion}Conclusion}
In summary, we have proposed an adaptive controller based on the
speed-gradient method, to tune the feedback gain of time-delayed
feedback control to an optimal value. We have shown that the
adaptation algorithm can find appropriate values for the feedback gain and thus
stabilize the desired periodic orbit or fixed point. This has been realized both for the stabilization of an unstable focus in a generic model and the
stabilization of an unstable periodic orbit embedded in a chaotic
attractor. We have demonstrated the robustness of our method to different
initial conditions and noise. We stress that this adaptive controller may especially be useful for systems with unknown or slowly changing
parameters where the domains of stability in parameter space are
unknown. In particular, we have shown by a simulation with a drifting bifurcation
parameter $\lambda$ that our method is able to follow such slow
parameter drifts. It should be noted that the automatic adjustment of the feedback gain
$K$ is possible without changing the value of the adaptation gain $\gamma$ of the speed-gradient method. This shows that the algorithm is robust and simple to apply. Our method might be used to tune more than one parameter, increasing its range of possible application. 

\begin{acknowledgements}
This work was supported by Deutsche Forschungsgemeinschaft in the
framework of SFB 910. J.~Lehnert, P.~H\"ovel and E.~Sch\"oll acknowledge the
support by the German-Russian Interdisciplinary Science Center (G-RISC) funded by the German Federal
Foreign Office via the German Academic Exchange Service
(DAAD). P. H\"ovel  acknowledges also support by the BMBF under the
grant no. 01GQ1001B. P.~Guzenko thanks the DAAD program "Michail Lomonosov (B)" for
the support of this work. A.~L.~Fradkov acknowledges support of Russian Federal Program "Cadres"
(goscontracts 16.740.11.0042, 14.740.11.0942) and RFBR (project 11-08-01218).

\end{acknowledgements}

\bibliographystyle{prsty-fullauthor}

\begin{thebibliography}{10}
\bibitem{OTT90}
E. Ott, C. Grebogi, and J.~A. Yorke, Phys.~Rev.~Lett. {\bf 64},  1196  (1990).

\bibitem{SCH07}
{\em Handbook of Chaos Control}, edited by E. Sch{\"o}ll and H.~G. Schuster
  (Wiley-VCH, Weinheim, 2008), second completely revised and enlarged edition.

\bibitem{PYR92}
K. Pyragas, Phys.~Lett.~A {\bf 170},  421  (1992).

\bibitem{BAL05}
A.~G. Balanov, N.~B. Janson, and E. Sch{\"o}ll, Phys.~Rev.~E {\bf 71},  016222
  (2005).

\bibitem{AHL04}
A. Ahlborn and U. Parlitz, Phys.~Rev.~Lett. {\bf 93},  264101  (2004).

\bibitem{ROS04a}
M.~G. Rosenblum and A.~S. Pikovsky, Phys.~Rev.~Lett. {\bf 92},  114102  (2004).

\bibitem{HOE05}
P. H{\"o}vel and E. Sch{\"o}ll, Phys.~Rev.~E {\bf 72},  046203  (2005).

\bibitem{SOC94}
J.~E.~S. Socolar, D.~W. Sukow, and D.~J. Gauthier, Phys.~Rev.~E {\bf 50},  3245
   (1994).

\bibitem{DAH07}
T. Dahms, P. H{\"o}vel, and E. Sch{\"o}ll, Phys.~Rev.~E {\bf 76},  056201
  (2007).

\bibitem{BLE96}
M.~E. Bleich and J.~E.~S. Socolar, Phys.~Lett.~A {\bf 210},  87  (1996).

\bibitem{JUS97}
W. Just, T. Bernard, M. Ostheimer, E. Reibold, and H. Benner, Phys.~Rev.~Lett.
  {\bf 78},  203  (1997).

\bibitem{JUS98}
W. Just, D. Reckwerth, J. M{\"o}ckel, E. Reibold, and H. Benner,
  Phys.~Rev.~Lett. {\bf 81},  562  (1998).

\bibitem{PYR01}
K. Pyragas, Phys.~Rev.~Lett. {\bf 86},  2265  (2001).

\bibitem{YAN06}
S. Yanchuk, M. Wolfrum, P. H{\"o}vel, and E. Sch{\"o}ll, Phys.~Rev.~E {\bf 74},
   026201  (2006).

\bibitem{HIN11}
R. Hinz, P. H{\"o}vel, and E. Sch{\"o}ll, Chaos {\bf 21},  023114  (2011).

\bibitem{BAB02}
N. Baba, A. Amann, E. Sch{\"o}ll, and W. Just, Phys.~Rev.~Lett. {\bf 89},
  074101  (2002).

\bibitem{NAK97}
H. Nakajima, Phys.~Lett.~A {\bf 232},  207  (1997).

\bibitem{FIE07}
B. Fiedler, V. Flunkert, M. Georgi, P. H{\"o}vel, and E. Sch{\"o}ll,
  Phys.~Rev.~Lett. {\bf 98},  114101  (2007).

\bibitem{JUS07}
W. Just, B. Fiedler, V. Flunkert, M. Georgi, P. H{\"o}vel, and E. Sch{\"o}ll,
  Phys.~Rev.~E {\bf 76},  026210  (2007).

\bibitem{FRA79}
A.~L. Fradkov, Autom. Remote Control {\bf 40},  1333  (1979).

\bibitem{FRA98a}
A.~L. Fradkov and A.~Y. Pogromsky, {\em {Introduction to Control of
  Oscillations and Chaos}} (World Scientific, Singapore, 1998).

\bibitem{FRA05b}
A.~L. Fradkov, Physics-Uspekhi {\bf 48},  103  (2005).

\bibitem{FRA07}
A.~L. Fradkov, {\em {Cybernetical Physics: From Control of Chaos to Quantum
  Control}} (Springer, Heidelberg, Germany, 2007).

\bibitem{GUZ08}
P.~Y. Guzenko, P. H{\"o}vel, V. Flunkert, A.~L. Fradkov, and E. Sch{\"o}ll,
  Adaptive Tuning of Feedback Gain in Time-Delayed Feedback Control, Proc. 6th
  EUROMECH Nonlinear Dynamics Conference (ENOC-2008), ed. A. Fradkov, B.
  Andrievsky, IPACS Open Access Library http://lib.physcon.ru (e-Library of the
  International Physics and Control Society), 2008.

\bibitem{FRA99a}
A.~L. Fradkov, I.~V. Miroshnik, and V.~O. Nikiforov, {\em Nonlinear and
  Adaptive Control of Complex Systems} (Kluwer, Dordrecht, 1999).

\bibitem{AST08a}
A. Astolfi, D. Karagiannis, and R. Ortega, {\em Nonlinear and Adaptive Control
  with Applications} (Springer, Heidelberg, 2008).

\bibitem{KRS95}
M. Krstic, I. Kanellakopoulos, and P. Kokotovic, {\em Nonlinear and Adaptive
  Control Design} (Wiley, New York, 1995).

\bibitem{BEC02}
O. Beck, A. Amann, E. Sch{\"o}ll, J.~E.~S. Socolar, and W. Just, Phys.~Rev.~E
  {\bf 66},  016213  (2002).

\bibitem{AST08}
Y.~A. Astrov, A.~L. Fradkov, and P.~Y. Guzenko, Phys. Rev. E {\bf 77},  026201
  (2008).

\bibitem{FLU11}
V. Flunkert and E. Sch{\"o}ll, Phys. Rev. E {\bf 84},  016214  (2011).

\end{thebibliography}

\end{document}